\documentclass[aip, amsmath, amssymb, reprint]{revtex4-1}

\usepackage{graphicx}
\usepackage{dcolumn}
\usepackage{bm}

\usepackage[utf8]{inputenc}
\usepackage[T1]{fontenc}
\usepackage{etoolbox}

\usepackage{todonotes}
\usepackage{cleveref}
\usepackage{siunitx}
\usepackage{eucal}
\usepackage{microtype}
\usepackage{newtxtext}
\usepackage{newtxmath}

\setuptodonotes{inline}

\usepackage{verbatim}

\definecolor{Red}{rgb}{1,0,0}

\begin{document}

\title{Josephson effect in a fractal geometry}
\author{Morten Amundsen}
    \email[Corresponding author: ]{morten.amundsen@ntnu.no}
    \affiliation{Center for Quantum Spintronics, Department of Physics,\\\; Norwegian University of Science and Technology, NO-7491 Trondheim, Norway}
\author{Vladimir Juri\v{c}i\'{c}}
    \affiliation{Departamento de F\'isica, Universidad T\'ecnica Federico Santa Mar\'ia, \\\; Casilla 110, Valpara\'iso, Chile}
    \affiliation{Nordita, KTH Royal Institute of Technology and Stockholm University, \\\; Hannes Alfvéns väg 12, SE-106 91 Stockholm, Sweden}
\author{Jabir Ali Ouassou}
    \affiliation{Department of Computer Science, Electrical Engineering and Mathematical Sciences,\\\; Western Norway University of Applied Sciences, NO-5528 Haugesund, Norway}

\begin{abstract}
 The Josephson effect is a hallmark signature of the superconducting state, which, however, has been sparsely explored in non-crystalline superconducting materials. Motivated by this,  we consider a Josephson junction consisting of two superconductors with a fractal metallic interlayer, which is patterned as a \emph{Sierpiński carpet} by removing atomic sites in a self-similar and scale-invariant manner. We here show that the fractal geometry has direct observable consequences on the Josephson effect. 
 In particular, we demonstrate that the  form of the supercurrent-magnetic field relation as the fractal generation number increases can be directly related to the self-similar fractal geometry of the normal metallic layer. Furthermore, the maxima  of the corresponding diffraction pattern directly encode the self-repeating fractal structure in the course of fractal generation, implying that  the corresponding magnetic length directly  probes the shortest length scale in the given fractal generation. Our results should motivate future experimental efforts to verify these predictions in designer quantum materials, and motivate future pursuits regarding fractal-based SQUID devices.
\end{abstract}

\maketitle
Fractals are paradigmatic non-crystalline structures featuring a non-integer spatial dimension with a self-similarly repeating pattern at smaller and smaller length scales, which are rather ubiquitous in Nature, from living organisms to geological objects.~\cite{mandelbrot1982fractal}
In the quantum realm, they are realized most commonly in terms of wavefunctions exhibiting a fractal behavior  when electrons confined on a plane are subjected to a perpendicular magnetic field, e.g.\ in the Hofstadter butterfly~\cite{Hofstadter1976}, as well as in disordered electronic systems.~\cite{Soukulis-1984,Schreiber-1991}
The interplay of fractal geometry in non-integer dimensions and  collective behavior of electrons and spins living on fractals~\cite{Gefen-PRL1980,gefen1984phase,Gefen-PRL1983} has attracted renewed interest  with advances in assembling of fractal structures~\cite{Shang2015} and the experimental observation of the emerging fractal features of the electronic wavefunctions.~\cite{kempkes2019}
These observations spurred further interest in this problem, especially in light  of possible nontrivial  topological properties  of the electronic wavefunctions on the fractals.~\cite{Neupert-2018,Prem-2019,Shengjun-2020,Manna-PRR-2020,Fremling-PRR-2020,Yang2020,Manna-PRA-2022,Manna-PRB-2022,Ivaki2022,Manna2023}
Quite surprisingly, however, superconductivity in this context has been rather sparsely explored.~\cite{Wang-1994,Ausloos1991,Bak2003,Hyung2005,Feigelman2010,Meng2024,Poccia2011, meyer_dimensional_2002, kingni_analysis_2019,manna2022noncrystalline}
Intriguingly, a very recent study has however suggested that artificial fractal geometries can enhance the critical temperature~$T_c$ of superconductors.\cite{iliasov2024} 

Superconductivity is a macroscopic quantum phenomenon, where a large number of electron pairs all condense into the same macroscopic quantum state.
This quantum state can be described using a macroscopic wave function $\Psi(\bm{r}, t)=|\Psi(\bm{r}, t)| e^{i\varphi(r,t)}$. When two different superconductors are connected to the same metal,  their wave functions can leak into the metal and interfere with each other.
This can result in the \emph{Josephson effect},\cite{josephson1962,golubov2004a} whereby a current that depends on the phase difference $\delta\varphi$ between the wavefunctions in the two superconductors flows through the normal metal. Josephson junctions have found numerous applications, ranging from precise metrology (e.g.\ the SQUID) to novel computing paradigms (e.g.\ RSFQ logic or phase qubits).

\begin{figure}[t!]
  \includegraphics[width=\columnwidth]{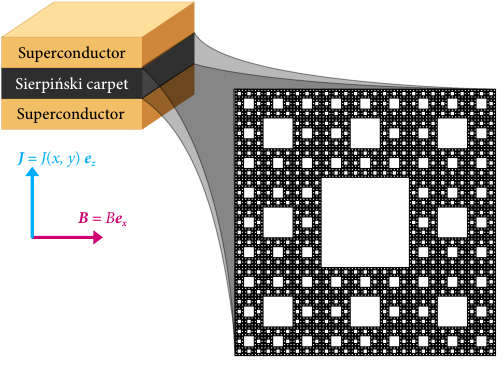}
  \caption{
    Geometry of  the S/N/S Josephson junction with the N layer exhibiting a  fractal (self-similar) structure. The setup consists of  two superconductors (yellow)  connected by a normal metal (black) that  has been patterned as a Sierpi\'nski fractal by recursively creating insulating holes according to \cref{eq:3}.
    A homogeneous in-plane magnetic field $\bm B \sim \bm{e}_x$ is then applied to the Josephson junction, causing spatially varying screening currents $\bm{J} \sim \bm{e}_z$ to flow between the superconductors.
    Depending on the magnitude $B$ of the applied magnetic field, the screening currents in different parts of the interlayer may either interfere constructively or destructively, causing a kind of interference pattern to appear in the current--field relation $I(B)$.
  }
  \label{fig:geometry}
\end{figure}

\begin{figure}[t!]
    \centering
    \includegraphics[width=\columnwidth]{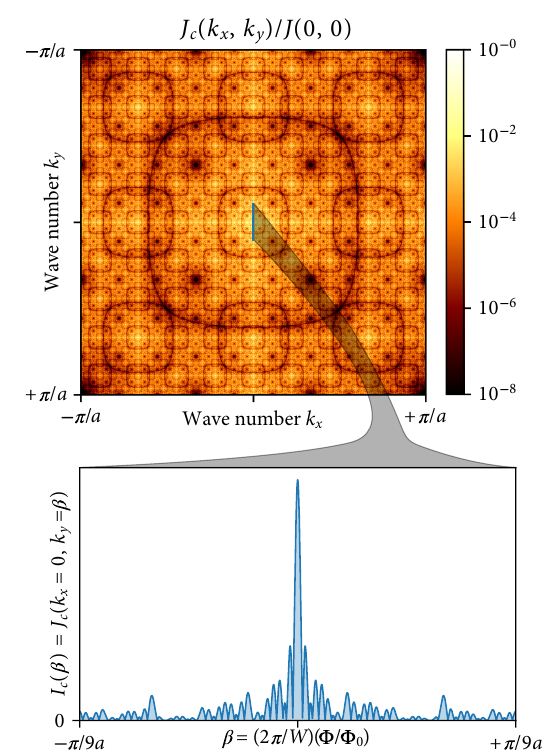}
    \caption{%
        Fourier transform $J_c(k_x, k_y)$ of the current distribution in a Sierpi\'nski Josephson junction (upper panel). We here  considered a 6th-order Sierpi\'nski lattice as the interlayer.
        According to the Dynes--Fulton method, the net current $I(\beta)$ that arises for an applied magnetic field $B \sim \beta$ along the $x$ axis is found from the Fourier-transformed current density along the $y$ axis, as indicated by the blue line in the upper panel, with the result shown in the lower panel.
    }
    \label{fig:fft}
\end{figure}

\begin{figure}[t!]
    \centering
    \includegraphics{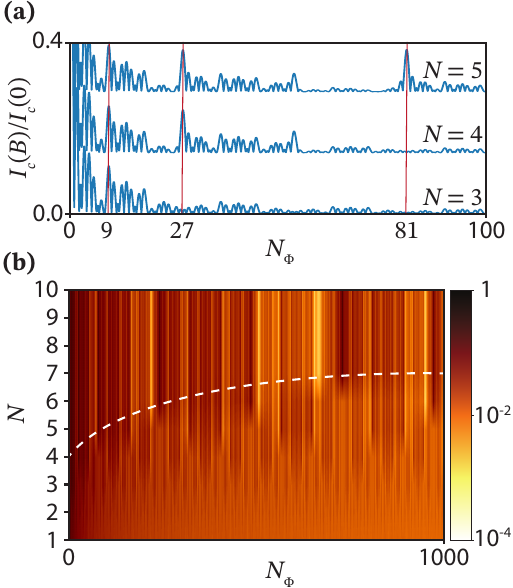}
    \caption{%
    The critical current $I_c$ as a function of magnetic field, $N_\Phi = BdW/\Phi_0$, for various generations $N$ of the Sierpi\'nski carpet. (a)  Line plots for $N\in\left\{3,4,5\right\}$, shifted vertically for clarity. The vertical lines indicate the location of new peaks introduced by an increase in $N$. (b) A surface plot of $I_c(B)$ for a wider range of values of $N$ and $N_\Phi$. The dashed line is a guide for the eye, showing approximately where $I_c(B)$ becomes independent of $N$.
    }
    \label{fig:diffpattern}
\end{figure}

Given the versatility of the Josephson effect and the recent advances in fractal geometries, there is ample opportunity for the discovery of novel applications by combining these two concepts. In spite of this, the Josephson effect in fractal geometries has  received only very limited attention in the literature so far.~\cite{meyer_dimensional_2002,kingni_analysis_2019,kruchinin_nonlinear_2005}
In this paper, we fill this gap by considering a heterostructure made of a sandwich of conventional $s$-wave superconductors~(S) and a normal material~(N) with a fractal geometry.
In particular, we show that the fractal geometry, here taken to be a Sierpi\'nski carpet (fig.~\ref{fig:geometry}), has direct observable consequences  on the Josephson effect. The form of the supercurrent-magnetic field relation in the process of the fractal generation, as we show,  can be directly related to the self-similar fractal pattern of the normal metallic layer, as shown in fig.~\ref{fig:fft}. Furthermore, the corresponding diffraction pattern, through its maxima,  encodes the self-repeating fractal structure in the course of fractal generation (fig.~\ref{fig:diffpattern}), which implies that  the corresponding magnetic length directly  probes the shortest length scale in the given fractal generation. Our findings should motivate experimental pursuits  to probe our theoretical predictions in the designer quantum materials.

We consider an S/N/S Josephson junction as illustrated in \cref{fig:geometry}.\cite{josephson1962}
We take the stack to be oriented along the $z$ axis, and to have dimensions $W \times W$ in the $xy$ plane.
Each S layer is a conventional BCS superconductor,\cite{bardeen1957} and is assumed to be much longer than the magnetic penetration depth~$\lambda$ along the $z$ direction.
The N layer is a thin layer of a nominally normal metal, but has in our case been patterned as a Sierpi\'nski carpet: It contains a fractal pattern of holes, where atoms have been removed from the otherwise normal metal in a self-similar and scale-invariant manner (see \cref{fig:geometry}).
In an experiment, it is likely easier to deposit a thin layer of an electric insulator at the locations of the holes in our geometry, which would have the same effect as our missing lattice sites.
Artificial fractal geometries in condensed matter systems---including Sierpi\'nski fractals---have in recent years been experimentally realized.\cite{kempkes2019}

The Josephson effect between the two superconductors produces dissipationless currents that flow through the fractal layer whenever a phase difference is enforced. We assume that the Josephson penetration depth is much larger than the in-plane dimensions of the system, and hence any self-screening effects may be neglected, meaning that the currents flow solely in the $z$ direction. 
The resulting current density therefore takes the form
\begin{equation}
    \bm{J}(x,y) = J(x,y)\,\bm{e}_z = J_0g(x,y) \sin[\varphi(x,y)]\,\bm{e}_z,
    \label{eq:crit-real}
\end{equation}
where $\delta\varphi$ is the gauge-invariant phase difference between two vertically separated points deep inside each superconductor, and $g(x,y)$ is a form factor, indicating the shape of the fractal.
To leading order, we have here assumed that the critical current density $J(x,y)$ is uniform and equal to $J_0$ at coordinates without a hole, whereas it drops to zero in the insulating holes. The critical current density is thus given as $J_c(x,y) = J_0g(x,y)$.
If we express the coordinates $x = ia$ and $y = ja$ in terms of a lattice constant~$a$, we can then write
\begin{equation}
    g(x, y) = S^N_{ij},
    \label{eq:2}
\end{equation}
where $S^N_{ij}$ are matrix elements of an $N$'th-order Sierpi\'nski carpet.
The Sierpi\'nski carpet is a fractal structure that has the properties of self-similarity and scale invariance---at least down to the scale of individual unit cells---and is illustrated in \cref{fig:geometry}.
Such a lattice may be constructed using the recursive algorithm
\begin{align}
  \label{eq:3}
  \bm{S}^{N} &= \bm{S}^{N-1} \otimes \bm{S}^{1},
  &
  \bm{S}^1 &= 
  \begin{pmatrix}
    \,1 & 1 & 1\; \\
    \,1 & 0 & 1\; \\
    \,1 & 1 & 1\;
  \end{pmatrix},
  &
  \bm{S}^0 &=
  1,
\end{align}
where $\otimes$ is the Kronecker matrix product.
This implies that we can also write the $N^{\rm th}$ order Sierpi\'nski carpet $\bm{S}^N = (\bm{S}^1)^{\otimes N}$ compactly as a Kronecker matrix power.

We now consider a magnetic field $\bm{B} = B\bm{e}_x$ applied in the thin-film plane of the fractal interlayer. This produces screening currents in the $yz$ plane of the superconductors, within a distance on the order of their penetration depth $\lambda$ away from the interface to the normal metal. These currents take the form of elongated loops that flow purely vertically through the thin normal metal, and give rise to a position-dependence in the gauge-invariant phase difference, $\phi = \phi_0 - \frac{2e}{\hbar} Bdy $, where $\phi_0$ is an externally enforced phase difference, and $d = \ell + 2\lambda$ is the effective thickness of the junction in the $z$ direction, with $\ell$ the thickness of the normal metal. In our analysis of the supercurrent through the junction, we have assumed that the effective thickness~$d$ is much smaller than the superconducting coherence length~$\xi$. The result are spatially varying current contributions in the normal metal that can interfere constructively or destructively depending on the precise value of the magnetic field~$B$. The total current flowing through the system at a given $B$ is given as the surface integral over the cross section, which is quadratic with widths $W$,
\begin{equation}
I(B,\phi_0) = J_0\int_{-W/2}^{W/2}\int_{-W/2}^{W/2} dx dy \; g(x,y)\sin\left[\phi_0 - \frac{2e}{\hbar} Bdy\right], 
\end{equation}
As was first observed by Dynes and Fulton,\cite{dynes1971a} this expression is simply the imaginary part of the Fourier transform of the current density, which in our case becomes the Fourier transform of the fractal form factor, $g(k_x,k_y)$. Furthermore, since $g(x,y)$ has inversion symmetry, $g(k_x,k_y)$ is purely real. The critical current is thus found at $\phi = \pi/2$, and is given as 
\begin{equation}
    I_c(B) = |J_c(k_x=0, k_y=\beta)| = J_0\int d\bm{r}\; g(\bm{r}) e^{-i\beta y},
    \label{eq:4}
\end{equation}
with $\bm{r} = x\bm{e}_x + y\bm{e}_y$. Here, the magnetic field $B$ is parametrized via $\beta = (2\pi/W) (\Phi / \Phi_0)$, where $\Phi = BWd$ is the net flux passing through the central parts of the Josephson junction, and $\Phi_0 = h/2e$ is the flux quantum.

Using the approach outlined above, the current--field relation~$I_c(B)$ can be extracted as follows.
First, an $N^{\rm th}$ order Sierpi\'nski carpet is constructed recursively using \cref{eq:3}.
This is subsequently employed to find  $J_c(x, y)$ according to \cref{eq:crit-real,eq:2}, which can then be run through a 2D Fast Fourier Transform (FFT) algorithm to obtain $J_c(k_x, k_y)$.
Finally, we can then simply extract the critical current along a line in the 2D plane, as determined by the direction of the applied field and the chosen gauge.
This procedure, and the resulting $I_c(B)$, are shown in \cref{fig:fft} for a junction with a 6th-order Sierpi\'nski interlayer.
Unsurprisingly, the Fourier transform of a fractal contains self-similar features also, and ends up having structures on all length scales from $\pi/W$ to $\pi/a$.
The resulting field dependence has a very different structure from the well-known Fraunhofer and SQUID patterns that arise in comparable non-fractal junctions.

We can arrive at this result by purely analytical means. This can be done by treating current as a parallel coupling of the current passing through each of the constituent elements of a generation of the Sierpi\'nski carpet. To illustrate this we consider, for instance, the first generation, as is obtained from \cref{fig:geometry} by retaining only the largest hole in the center. In that case, the total current may be split into three parts, two of which form rectangular junctions with dimensions $W\times W/3$, and one part which has the form of a SQUID formed by two junctions with dimensions $W/3 \times W/3$ and an equally sized hole in between. The rectangular contributions provide a current in a Fraunhofer pattern, $J_0 \frac{W^2}{3} \mathrm{sinc}\left( \pi N_\Phi\right)$, whereas the SQUID contribution gives a current $2J_0 \left(\frac{W}{3}\right)^2\mathrm{sinc}\left(\frac{\pi N_\Phi}{3}\right)\cos\left(\frac{2\pi N_\Phi}{3}\right)$. Here, $N_\Phi = BdW/\Phi_0$ is the number of flux quanta passing through the system, and $\mathrm{sinc}(x) = \sin(x)/x$. Adding up each contribution gives the critical current for the first-generation Sierpi\'nski carpet
\begin{equation*}
I^{(1)}_c(B) = \frac{J_0 A_F^{(1)}}{4}\left|\mathrm{sinc}\left(\frac{\pi N_\Phi}{3}\right)\left[1 + 3\cos\frac{2\pi N_\Phi}{3}\right]\right|, 
\end{equation*}
where $A_F^{(N)} = \left(\frac{8}{9}\right)^n W^2$ is the surface area of the fractal at generation $n$. An expression for arbitrary generation $N$ can be obtained by repeating this process recursively, replacing each uniform sub-square with the shape of the previous generation. This leads to
\begin{equation}
    I^{(N)}_c(B) = \frac{J_0 A^{(N)}_F}{4^N} \left|\, \mathrm{sinc}\left(\frac{\pi N_\Phi}{3^N} \right) \prod_{n=1}^N \left[ 1 + 3 \cos\left(\frac{2\pi N_\Phi}{3^n} \right) \right] \,\right|,
    \label{eq:Ianalytical}
\end{equation}
The 0th-order Sierpi\'nski carpet is equivalent to a uniform square lattice, and indeed we regain the conventional Fraunhofer pattern for $N=0$ when no terms appear in the right-hand product, $I^{(0)}_c(B) = J_0W^2\mathrm{sinc}(\pi N_\Phi)$.
In the opposite limit $N \to \infty$ of a high-order Sierpi\'nski carpet, the prefactor approaches $\mathrm{sinc}(0) = 1$.
The current--field relation then reduces to a  product of similar factors $1 + 3\cos(\beta W_n)$ arising from the Sierpi\'nski pattern on different scales $W_n = W/3^n$.
This result shows clearly that $I_c(B)$ exhibits self-similarity across different length scales, which arises from the fractal geometry of the Josephson junction interlayer.
This self-similarity can be seen in the 2D Fourier transform in \cref{fig:fft}, where it manifests as superimposed lattices of $3^n \times 3^n$ squares.  

The presence of multiple length scales has additional nontrivial observable consequences, such as the sensitivity of the  diffraction pattern on the order of fractal generation. In \cref{fig:diffpattern}(a) the diffraction pattern is shown for a selection of fractal generations, $N\in\left\{3,4,5\right\}$. It can be seen that each new generation introduces new features in the diffraction pattern which were not present in the previous one. These features appear in limited ranges of $N_\Phi$, and are characterized by a single peak of constant height, the location of which is indicated by the vertical lines in \cref{fig:diffpattern}(a). Indeed, for a fixed $N$, we observe that whenever $N_\Phi = 3^q$ for an integer $q$, the square brackets in \cref{eq:Ianalytical} become constant and maximal for all $n \leq q$. For $q < N$, $I_c^{(N)}$ increases with $q$, whereas for $q \geq N$ it vanishes. The result is a local maximum at $q = N-1$, at which point the critical current becomes $I_c^{(N)} = 3\sqrt{3}J_0A_F^{(N)} / 16\pi$. The peak occurring at $N_\Phi = 3^{N-1}$ is a manifestation of the fact that the magnetic length scale becomes small enough to resolve the lowest length scales of the fractal. Alternatively, one may say that when  fractal generation is increased from $N$ to $N+1$, the smallest length scale is reduced by a factor of 3. Hence, a flux density corresponding to a rescaled magnetic field of $3B$ in generation $N+1$, produces a similar response as $B$ in generation $N$. Another interesting and related feature is that the critical current for any $N_\Phi$ eventually becomes independent of $N$. This is demonstrated in \cref{fig:diffpattern}(b), which shows a surface plot of the critical current as a function of $N$ and $N_\Phi$. The dashed line indicates approximately where this saturation occurs, as determined by visual inspection. For $N = 7$, the diffraction pattern remains constant at least for $N_\Phi \leq 1000$. This behavior is once again a result of the underlying fractal structure,  namely, as $N$ increases, smaller and smaller length scales are introduced, which eventually become smaller than the magnetic length scale. Their features cannot be resolved by the applied flux, and so the system remains uninfluenced.

The confluence between the fractal geometry and the Josephson effect discussed in our work may be instrumental to developing SQUID devices with an increased magnetic flux resolution and larger operating range due to the logarithmic scale of the response. This feature is directly implied by the sensitivity of the diffraction pattern on the order of the fractal generation. As shown in Fig.~\ref{fig:diffpattern},  the system exhibits self-similar features in the critical current $I_c(B)$. Notably, there is a significant spike at flux values $N_\Phi = 3^n$, where $n < N$ are integers less than the fractal generation $N$.
Consequently, such a fractal structure might naturally lead to a logarithmic scale response in terms of the magnetic flux  in such a device.
In turn,  a rather increased sensitivity of the fractal-based SQUID devices is expected in comparison with the conventional ones. More precise characterization of the possible practical applications of our results is, however, beyond the scope of the present work.%

Here, we have focused on the specific case where the magnetic field $\bm{B} \sim B\bm{e}_x$ is applied along the $x$ axis.
Note however that a similar Dynes--Fulton analysis is applicable for a magnetic field $\bm{B}(\theta) = B (\cos \theta \, \bm{e}_x + \sin \theta \, \bm{e}_y)$ oriented along any direction in the $x-y$ plane, where each such direction would provide a slightly different current--field relation $I_c(B \,|\, \theta)$.
In principle, one could in an experiment use the $I_c(B \,|\, \theta)$ curves obtained for different field directions $\theta$ to reconstruct the 2D Fourier transform $J_c(k_x, k_y)$ of the current density, and thus to obtain the exact current distribution $J_c(x, y)$ in the interlayer.
Such an approach might be particularly interesting in Josephson junctions where the interlayer might not be an artificially constructed fractal as considered here, but rather a naturally occurring fractal or quasicrystal.
In that case, this kind of Dynes--Fulton analysis might provide a way to validate and elucidate the non-crystalline,  fractal or quasicrystalline, structure of the interlayer.

In this paper, we have considered a fractal Josephson junction composed of two superconductors separated by a Sierpi\'nski-patterned normal metal.
We have analytically calculated the current--field relation $I_c(B)$ that arises in such junctions, and shown that this experimental signature retains a crucial property of fractals: self-similarity across length scales. In particular, the obtained  diffraction pattern is directly related to the self-similar fractal geometry probed  through the magnetic length. Such an approach is valid at sufficiently long length scales, corresponding to the
coarse features in a Sierpinski fractal. However, 
when  near-atomic length scales are probed, the existence of finite-size
effects necessitates a more detailed approach to get more detailed, quantitative predictions. Finally, we note that while each fractal is associated with a Hausdorff dimension (which for the Sierpinski carpet is given as $\ln 8 / \ln 3$), this quantity is not unique to a particular fractal. Hence, it is by itself not sufficient to predict our diffraction pattern.

Our results should be consequential for experiments, particularly in the designer quantum materials, where the realization of the normal fractal metallic layer should be feasible. Notably, we propose that fractal Josephson junctions may be useful as a SQUID-like device with an intrinsic logarithmic-scale response, which thus might be useful over a larger operating range than conventional devices. This can be considered a superconducting analogue to e.g.\ the established fractal antennae and resonators~\cite{fractal_antennae,Cohen_fractal,Werner_fractal}, where the fractal structure extends the range of frequencies at which the device can operate with fidelity.

\noindent%
M.A. was supported by the Research Council of Norway through its Centres of Excellence funding scheme Grant No. 262633 ``QuSpin''. V.J. acknowledges the support of  the Swedish Research Council Grant No. VR 2019-04735 and  Fondecyt (Chile) Grant No. 1230933. 
\newline
\textbf{Data Availability}\\
\noindent
Data sharing is not applicable to this article as no new data were created or analyzed in this study.

\nocite{*}
\bibliography{references}

\providecommand{\noopsort}[1]{}\providecommand{\singleletter}[1]{#1}%
\begin{thebibliography}{39}%
\makeatletter
\providecommand \@ifxundefined [1]{%
 \@ifx{#1\undefined}
}%
\providecommand \@ifnum [1]{%
 \ifnum #1\expandafter \@firstoftwo
 \else \expandafter \@secondoftwo
 \fi
}%
\providecommand \@ifx [1]{%
 \ifx #1\expandafter \@firstoftwo
 \else \expandafter \@secondoftwo
 \fi
}%
\providecommand \natexlab [1]{#1}%
\providecommand \enquote  [1]{``#1''}%
\providecommand \bibnamefont  [1]{#1}%
\providecommand \bibfnamefont [1]{#1}%
\providecommand \citenamefont [1]{#1}%
\providecommand \href@noop [0]{\@secondoftwo}%
\providecommand \href [0]{\begingroup \@sanitize@url \@href}%
\providecommand \@href[1]{\@@startlink{#1}\@@href}%
\providecommand \@@href[1]{\endgroup#1\@@endlink}%
\providecommand \@sanitize@url [0]{\catcode `\\12\catcode `\$12\catcode
  `\&12\catcode `\#12\catcode `\^12\catcode `\_12\catcode `\%12\relax}%
\providecommand \@@startlink[1]{}%
\providecommand \@@endlink[0]{}%
\providecommand \url  [0]{\begingroup\@sanitize@url \@url }%
\providecommand \@url [1]{\endgroup\@href {#1}{\urlprefix }}%
\providecommand \urlprefix  [0]{URL }%
\providecommand \Eprint [0]{\href }%
\providecommand \doibase [0]{http://dx.doi.org/}%
\providecommand \selectlanguage [0]{\@gobble}%
\providecommand \bibinfo  [0]{\@secondoftwo}%
\providecommand \bibfield  [0]{\@secondoftwo}%
\providecommand \translation [1]{[#1]}%
\providecommand \BibitemOpen [0]{}%
\providecommand \bibitemStop [0]{}%
\providecommand \bibitemNoStop [0]{.\EOS\space}%
\providecommand \EOS [0]{\spacefactor3000\relax}%
\providecommand \BibitemShut  [1]{\csname bibitem#1\endcsname}%
\let\auto@bib@innerbib\@empty
\bibitem [{\citenamefont {Mandelbrot}(1982)}]{mandelbrot1982fractal}%
  \BibitemOpen
  \bibfield  {author} {\bibinfo {author} {\bibfnamefont {B.~B.}\ \bibnamefont
  {Mandelbrot}},\ }\href@noop {} {\emph {\bibinfo {title} {The fractal geometry
  of {Nature}}}}\ (\bibinfo  {publisher} {W.H. Freeman and Co.},\ \bibinfo
  {year} {1982})\BibitemShut {NoStop}%
\bibitem [{\citenamefont {Hofstadter}(1976)}]{Hofstadter1976}%
  \BibitemOpen
  \bibfield  {author} {\bibinfo {author} {\bibfnamefont {D.~R.}\ \bibnamefont
  {Hofstadter}},\ }\bibfield  {title} {\enquote {\bibinfo {title} {Energy
  levels and wave functions of {Bloch} electrons in rational and irrational
  magnetic fields},}\ }\href {\doibase 10.1103/PhysRevB.14.2239} {\bibfield
  {journal} {\bibinfo  {journal} {Phys. Rev. B}\ }\textbf {\bibinfo {volume}
  {14}},\ \bibinfo {pages} {2239--2249} (\bibinfo {year} {1976})}\BibitemShut
  {NoStop}%
\bibitem [{\citenamefont {Soukoulis}\ and\ \citenamefont
  {Economou}(1984)}]{Soukulis-1984}%
  \BibitemOpen
  \bibfield  {author} {\bibinfo {author} {\bibfnamefont {C.~M.}\ \bibnamefont
  {Soukoulis}}\ and\ \bibinfo {author} {\bibfnamefont {E.~N.}\ \bibnamefont
  {Economou}},\ }\bibfield  {title} {\enquote {\bibinfo {title} {Fractal
  character of eigenstates in disordered systems},}\ }\href {\doibase
  10.1103/PhysRevLett.52.565} {\bibfield  {journal} {\bibinfo  {journal}
  {Physical Review Letters}\ }\textbf {\bibinfo {volume} {52}},\ \bibinfo
  {pages} {565--568} (\bibinfo {year} {1984})}\BibitemShut {NoStop}%
\bibitem [{\citenamefont {Schreiber}\ and\ \citenamefont
  {Grussbach}(1991)}]{Schreiber-1991}%
  \BibitemOpen
  \bibfield  {author} {\bibinfo {author} {\bibfnamefont {M.}~\bibnamefont
  {Schreiber}}\ and\ \bibinfo {author} {\bibfnamefont {H.}~\bibnamefont
  {Grussbach}},\ }\bibfield  {title} {\enquote {\bibinfo {title} {Multifractal
  wave functions at the {Anderson} transition},}\ }\href {\doibase
  10.1103/PhysRevLett.67.607} {\bibfield  {journal} {\bibinfo  {journal}
  {Physical Review Letters}\ }\textbf {\bibinfo {volume} {67}},\ \bibinfo
  {pages} {607--610} (\bibinfo {year} {1991})}\BibitemShut {NoStop}%
\bibitem [{\citenamefont {Gefen}, \citenamefont {Mandelbrot},\ and\
  \citenamefont {Aharony}(1980)}]{Gefen-PRL1980}%
  \BibitemOpen
  \bibfield  {author} {\bibinfo {author} {\bibfnamefont {Y.}~\bibnamefont
  {Gefen}}, \bibinfo {author} {\bibfnamefont {B.~B.}\ \bibnamefont
  {Mandelbrot}}, \ and\ \bibinfo {author} {\bibfnamefont {A.}~\bibnamefont
  {Aharony}},\ }\bibfield  {title} {\enquote {\bibinfo {title} {Critical
  phenomena on fractal lattices},}\ }\href {\doibase
  10.1103/PhysRevLett.45.855} {\bibfield  {journal} {\bibinfo  {journal} {Phys.
  Rev. Lett.}\ }\textbf {\bibinfo {volume} {45}},\ \bibinfo {pages} {855--858}
  (\bibinfo {year} {1980})}\BibitemShut {NoStop}%
\bibitem [{\citenamefont {Gefen}\ \emph {et~al.}(1984)\citenamefont {Gefen},
  \citenamefont {Aharony}, \citenamefont {Shapir},\ and\ \citenamefont
  {Mandelbrot}}]{gefen1984phase}%
  \BibitemOpen
  \bibfield  {author} {\bibinfo {author} {\bibfnamefont {Y.}~\bibnamefont
  {Gefen}}, \bibinfo {author} {\bibfnamefont {A.}~\bibnamefont {Aharony}},
  \bibinfo {author} {\bibfnamefont {Y.}~\bibnamefont {Shapir}}, \ and\ \bibinfo
  {author} {\bibfnamefont {B.~B.}\ \bibnamefont {Mandelbrot}},\ }\bibfield
  {title} {\enquote {\bibinfo {title} {Phase transitions on fractals. {II.}
  {Sierpinski} gaskets},}\ }\href@noop {} {\bibfield  {journal} {\bibinfo
  {journal} {Journal of Physics A: Mathematical and General}\ }\textbf
  {\bibinfo {volume} {17}},\ \bibinfo {pages} {435} (\bibinfo {year}
  {1984})}\BibitemShut {NoStop}%
\bibitem [{\citenamefont {Gefen}, \citenamefont {Aharony},\ and\ \citenamefont
  {Alexander}(1983)}]{Gefen-PRL1983}%
  \BibitemOpen
  \bibfield  {author} {\bibinfo {author} {\bibfnamefont {Y.}~\bibnamefont
  {Gefen}}, \bibinfo {author} {\bibfnamefont {A.}~\bibnamefont {Aharony}}, \
  and\ \bibinfo {author} {\bibfnamefont {S.}~\bibnamefont {Alexander}},\
  }\bibfield  {title} {\enquote {\bibinfo {title} {Anomalous diffusion on
  percolating clusters},}\ }\href {\doibase 10.1103/PhysRevLett.50.77}
  {\bibfield  {journal} {\bibinfo  {journal} {Phys. Rev. Lett.}\ }\textbf
  {\bibinfo {volume} {50}},\ \bibinfo {pages} {77--80} (\bibinfo {year}
  {1983})}\BibitemShut {NoStop}%
\bibitem [{\citenamefont {Shang}\ \emph {et~al.}(2015)\citenamefont {Shang},
  \citenamefont {Wang}, \citenamefont {Chen}, \citenamefont {Dai},
  \citenamefont {Zhou}, \citenamefont {Kuttner}, \citenamefont {Hilt},
  \citenamefont {Shao}, \citenamefont {Gottfried},\ and\ \citenamefont
  {Wu}}]{Shang2015}%
  \BibitemOpen
  \bibfield  {author} {\bibinfo {author} {\bibfnamefont {J.}~\bibnamefont
  {Shang}}, \bibinfo {author} {\bibfnamefont {Y.}~\bibnamefont {Wang}},
  \bibinfo {author} {\bibfnamefont {M.}~\bibnamefont {Chen}}, \bibinfo {author}
  {\bibfnamefont {J.}~\bibnamefont {Dai}}, \bibinfo {author} {\bibfnamefont
  {X.}~\bibnamefont {Zhou}}, \bibinfo {author} {\bibfnamefont {J.}~\bibnamefont
  {Kuttner}}, \bibinfo {author} {\bibfnamefont {G.}~\bibnamefont {Hilt}},
  \bibinfo {author} {\bibfnamefont {X.}~\bibnamefont {Shao}}, \bibinfo {author}
  {\bibfnamefont {J.~M.}\ \bibnamefont {Gottfried}}, \ and\ \bibinfo {author}
  {\bibfnamefont {K.}~\bibnamefont {Wu}},\ }\bibfield  {title} {\enquote
  {\bibinfo {title} {Assembling molecular {Sierpi\'{n}ski} triangle
  fractals},}\ }\href {\doibase 10.1038/nchem.2211} {\bibfield  {journal}
  {\bibinfo  {journal} {Nature Chemistry}\ }\textbf {\bibinfo {volume} {7}},\
  \bibinfo {pages} {389--393} (\bibinfo {year} {2015})}\BibitemShut {NoStop}%
\bibitem [{\citenamefont {Kempkes}\ \emph {et~al.}(2019)\citenamefont
  {Kempkes}, \citenamefont {Slot}, \citenamefont {Freeney}, \citenamefont
  {Zevenhuizen}, \citenamefont {Vanmaekelbergh}, \citenamefont {Swart},\ and\
  \citenamefont {Smith}}]{kempkes2019}%
  \BibitemOpen
  \bibfield  {author} {\bibinfo {author} {\bibfnamefont {S.~N.}\ \bibnamefont
  {Kempkes}}, \bibinfo {author} {\bibfnamefont {M.~R.}\ \bibnamefont {Slot}},
  \bibinfo {author} {\bibfnamefont {S.~E.}\ \bibnamefont {Freeney}}, \bibinfo
  {author} {\bibfnamefont {S.~J.~M.}\ \bibnamefont {Zevenhuizen}}, \bibinfo
  {author} {\bibfnamefont {D.}~\bibnamefont {Vanmaekelbergh}}, \bibinfo
  {author} {\bibfnamefont {I.}~\bibnamefont {Swart}}, \ and\ \bibinfo {author}
  {\bibfnamefont {C.~M.}\ \bibnamefont {Smith}},\ }\bibfield  {title} {\enquote
  {\bibinfo {title} {Design and characterization of electrons in a fractal
  geometry},}\ }\href {\doibase 10.1038/s41567-018-0328-0} {\bibfield
  {journal} {\bibinfo  {journal} {Nature Physics}\ }\textbf {\bibinfo {volume}
  {15}},\ \bibinfo {pages} {127--131} (\bibinfo {year} {2019})}\BibitemShut
  {NoStop}%
\bibitem [{\citenamefont {Brzezi\'{n}ska}, \citenamefont {Cook},\ and\
  \citenamefont {Neupert}(2018)}]{Neupert-2018}%
  \BibitemOpen
  \bibfield  {author} {\bibinfo {author} {\bibfnamefont {M.}~\bibnamefont
  {Brzezi\'{n}ska}}, \bibinfo {author} {\bibfnamefont {A.~M.}\ \bibnamefont
  {Cook}}, \ and\ \bibinfo {author} {\bibfnamefont {T.}~\bibnamefont
  {Neupert}},\ }\bibfield  {title} {\enquote {\bibinfo {title} {Topology in the
  {Sierpi\'{n}ski}--{Hofstadter} problem},}\ }\href {\doibase
  10.1103/PhysRevB.98.205116} {\bibfield  {journal} {\bibinfo  {journal}
  {Physical Review B}\ }\textbf {\bibinfo {volume} {98}},\ \bibinfo {pages}
  {205116} (\bibinfo {year} {2018})}\BibitemShut {NoStop}%
\bibitem [{\citenamefont {Pai}\ and\ \citenamefont {Prem}(2019)}]{Prem-2019}%
  \BibitemOpen
  \bibfield  {author} {\bibinfo {author} {\bibfnamefont {S.}~\bibnamefont
  {Pai}}\ and\ \bibinfo {author} {\bibfnamefont {A.}~\bibnamefont {Prem}},\
  }\bibfield  {title} {\enquote {\bibinfo {title} {Topological states on
  fractal lattices},}\ }\href {\doibase 10.1103/PhysRevB.100.155135} {\bibfield
   {journal} {\bibinfo  {journal} {Physical Review B}\ }\textbf {\bibinfo
  {volume} {100}},\ \bibinfo {pages} {155135} (\bibinfo {year}
  {2019})}\BibitemShut {NoStop}%
\bibitem [{\citenamefont {Iliasov}, \citenamefont {Katsnelson},\ and\
  \citenamefont {Yuan}(2020)}]{Shengjun-2020}%
  \BibitemOpen
  \bibfield  {author} {\bibinfo {author} {\bibfnamefont {A.~A.}\ \bibnamefont
  {Iliasov}}, \bibinfo {author} {\bibfnamefont {M.~I.}\ \bibnamefont
  {Katsnelson}}, \ and\ \bibinfo {author} {\bibfnamefont {S.}~\bibnamefont
  {Yuan}},\ }\bibfield  {title} {\enquote {\bibinfo {title} {{Hall}
  conductivity of a {Sierpi\'{n}ski} carpet},}\ }\href {\doibase
  10.1103/PhysRevB.101.045413} {\bibfield  {journal} {\bibinfo  {journal}
  {Physical Review B}\ }\textbf {\bibinfo {volume} {101}},\ \bibinfo {pages}
  {045413} (\bibinfo {year} {2020})}\BibitemShut {NoStop}%
\bibitem [{\citenamefont {Manna}\ \emph {et~al.}(2020)\citenamefont {Manna},
  \citenamefont {Pal}, \citenamefont {Wang},\ and\ \citenamefont
  {Nielsen}}]{Manna-PRR-2020}%
  \BibitemOpen
  \bibfield  {author} {\bibinfo {author} {\bibfnamefont {S.}~\bibnamefont
  {Manna}}, \bibinfo {author} {\bibfnamefont {B.}~\bibnamefont {Pal}}, \bibinfo
  {author} {\bibfnamefont {W.}~\bibnamefont {Wang}}, \ and\ \bibinfo {author}
  {\bibfnamefont {A.~E.~B.}\ \bibnamefont {Nielsen}},\ }\bibfield  {title}
  {\enquote {\bibinfo {title} {Anyons and fractional quantum {Hall} effect in
  fractal dimensions},}\ }\href {\doibase 10.1103/PhysRevResearch.2.023401}
  {\bibfield  {journal} {\bibinfo  {journal} {Physical Review Research}\
  }\textbf {\bibinfo {volume} {2}},\ \bibinfo {pages} {023401} (\bibinfo {year}
  {2020})}\BibitemShut {NoStop}%
\bibitem [{\citenamefont {Fremling}\ \emph {et~al.}(2020)\citenamefont
  {Fremling}, \citenamefont {van Hooft}, \citenamefont {Smith},\ and\
  \citenamefont {Fritz}}]{Fremling-PRR-2020}%
  \BibitemOpen
  \bibfield  {author} {\bibinfo {author} {\bibfnamefont {M.}~\bibnamefont
  {Fremling}}, \bibinfo {author} {\bibfnamefont {M.}~\bibnamefont {van Hooft}},
  \bibinfo {author} {\bibfnamefont {C.~M.}\ \bibnamefont {Smith}}, \ and\
  \bibinfo {author} {\bibfnamefont {L.}~\bibnamefont {Fritz}},\ }\bibfield
  {title} {\enquote {\bibinfo {title} {Existence of robust edge currents in
  {Sierpi\'{n}ski} fractals},}\ }\href {\doibase
  10.1103/PhysRevResearch.2.013044} {\bibfield  {journal} {\bibinfo  {journal}
  {Physical Review Research}\ }\textbf {\bibinfo {volume} {2}},\ \bibinfo
  {pages} {013044} (\bibinfo {year} {2020})}\BibitemShut {NoStop}%
\bibitem [{\citenamefont {Yang}\ \emph {et~al.}(2020)\citenamefont {Yang},
  \citenamefont {Lustig}, \citenamefont {Lumer},\ and\ \citenamefont
  {Segev}}]{Yang2020}%
  \BibitemOpen
  \bibfield  {author} {\bibinfo {author} {\bibfnamefont {Z.}~\bibnamefont
  {Yang}}, \bibinfo {author} {\bibfnamefont {E.}~\bibnamefont {Lustig}},
  \bibinfo {author} {\bibfnamefont {Y.}~\bibnamefont {Lumer}}, \ and\ \bibinfo
  {author} {\bibfnamefont {M.}~\bibnamefont {Segev}},\ }\bibfield  {title}
  {\enquote {\bibinfo {title} {Photonic {Floquet} topological insulators in a
  fractal lattice},}\ }\href {\doibase 10.1038/s41377-020-00354-z} {\bibfield
  {journal} {\bibinfo  {journal} {Light: Science {\&} Applications}\ }\textbf
  {\bibinfo {volume} {9}},\ \bibinfo {pages} {128} (\bibinfo {year}
  {2020})}\BibitemShut {NoStop}%
\bibitem [{\citenamefont {Manna}\ \emph {et~al.}(2022)\citenamefont {Manna},
  \citenamefont {Duncan}, \citenamefont {Weidner}, \citenamefont {Sherson},\
  and\ \citenamefont {Nielsen}}]{Manna-PRA-2022}%
  \BibitemOpen
  \bibfield  {author} {\bibinfo {author} {\bibfnamefont {S.}~\bibnamefont
  {Manna}}, \bibinfo {author} {\bibfnamefont {C.~W.}\ \bibnamefont {Duncan}},
  \bibinfo {author} {\bibfnamefont {C.~A.}\ \bibnamefont {Weidner}}, \bibinfo
  {author} {\bibfnamefont {J.~F.}\ \bibnamefont {Sherson}}, \ and\ \bibinfo
  {author} {\bibfnamefont {A.~E.~B.}\ \bibnamefont {Nielsen}},\ }\bibfield
  {title} {\enquote {\bibinfo {title} {Anyon braiding on a fractal lattice with
  a local {Hamiltonian}},}\ }\href {\doibase 10.1103/PhysRevA.105.L021302}
  {\bibfield  {journal} {\bibinfo  {journal} {Physical Review A}\ }\textbf
  {\bibinfo {volume} {105}},\ \bibinfo {pages} {L021302} (\bibinfo {year}
  {2022})}\BibitemShut {NoStop}%
\bibitem [{\citenamefont {Manna}, \citenamefont {Nandy},\ and\ \citenamefont
  {Roy}(2022)}]{Manna-PRB-2022}%
  \BibitemOpen
  \bibfield  {author} {\bibinfo {author} {\bibfnamefont {S.}~\bibnamefont
  {Manna}}, \bibinfo {author} {\bibfnamefont {S.}~\bibnamefont {Nandy}}, \ and\
  \bibinfo {author} {\bibfnamefont {B.}~\bibnamefont {Roy}},\ }\bibfield
  {title} {\enquote {\bibinfo {title} {Higher-order topological phases on
  fractal lattices},}\ }\href {\doibase 10.1103/PhysRevB.105.L201301}
  {\bibfield  {journal} {\bibinfo  {journal} {Physical Review B}\ }\textbf
  {\bibinfo {volume} {105}},\ \bibinfo {pages} {L201301} (\bibinfo {year}
  {2022})}\BibitemShut {NoStop}%
\bibitem [{\citenamefont {Ivaki}\ \emph {et~al.}(2022)\citenamefont {Ivaki},
  \citenamefont {Sahlberg}, \citenamefont {P{\"o}yh{\"o}nen},\ and\
  \citenamefont {Ojanen}}]{Ivaki2022}%
  \BibitemOpen
  \bibfield  {author} {\bibinfo {author} {\bibfnamefont {M.~N.}\ \bibnamefont
  {Ivaki}}, \bibinfo {author} {\bibfnamefont {I.}~\bibnamefont {Sahlberg}},
  \bibinfo {author} {\bibfnamefont {K.}~\bibnamefont {P{\"o}yh{\"o}nen}}, \
  and\ \bibinfo {author} {\bibfnamefont {T.}~\bibnamefont {Ojanen}},\
  }\bibfield  {title} {\enquote {\bibinfo {title} {Topological random
  fractals},}\ }\href {\doibase 10.1038/s42005-022-01101-z} {\bibfield
  {journal} {\bibinfo  {journal} {Communications Physics}\ }\textbf {\bibinfo
  {volume} {5}},\ \bibinfo {pages} {327} (\bibinfo {year} {2022})}\BibitemShut
  {NoStop}%
\bibitem [{\citenamefont {Manna}\ and\ \citenamefont {Roy}(2023)}]{Manna2023}%
  \BibitemOpen
  \bibfield  {author} {\bibinfo {author} {\bibfnamefont {S.}~\bibnamefont
  {Manna}}\ and\ \bibinfo {author} {\bibfnamefont {B.}~\bibnamefont {Roy}},\
  }\bibfield  {title} {\enquote {\bibinfo {title} {Inner skin effects on
  non-{Hermitian} topological fractals},}\ }\href {\doibase
  10.1038/s42005-023-01130-2} {\bibfield  {journal} {\bibinfo  {journal}
  {Communications Physics}\ }\textbf {\bibinfo {volume} {6}},\ \bibinfo {pages}
  {10} (\bibinfo {year} {2023})}\BibitemShut {NoStop}%
\bibitem [{\citenamefont {Wang}\ \emph {et~al.}(1994)\citenamefont {Wang},
  \citenamefont {Li}, \citenamefont {Jiang}, \citenamefont {Zhang},\ and\
  \citenamefont {Tian}}]{Wang-1994}%
  \BibitemOpen
  \bibfield  {author} {\bibinfo {author} {\bibfnamefont {X.-B.}\ \bibnamefont
  {Wang}}, \bibinfo {author} {\bibfnamefont {J.-X.}\ \bibnamefont {Li}},
  \bibinfo {author} {\bibfnamefont {Q.}~\bibnamefont {Jiang}}, \bibinfo
  {author} {\bibfnamefont {Z.-H.}\ \bibnamefont {Zhang}}, \ and\ \bibinfo
  {author} {\bibfnamefont {D.-C.}\ \bibnamefont {Tian}},\ }\bibfield  {title}
  {\enquote {\bibinfo {title} {Effect of fractons in superconductors with
  fractal structure},}\ }\href {\doibase 10.1103/PhysRevB.49.9778} {\bibfield
  {journal} {\bibinfo  {journal} {Phys. Rev. B}\ }\textbf {\bibinfo {volume}
  {49}},\ \bibinfo {pages} {9778--9781} (\bibinfo {year} {1994})}\BibitemShut
  {NoStop}%
\bibitem [{\citenamefont {Ausloos}\ \emph {et~al.}(1991)\citenamefont
  {Ausloos}, \citenamefont {Laurent}, \citenamefont {Patapis}, \citenamefont
  {Politis}, \citenamefont {Luo}, \citenamefont {Godelaine}, \citenamefont
  {Gillet}, \citenamefont {Dang},\ and\ \citenamefont {Cloots}}]{Ausloos1991}%
  \BibitemOpen
  \bibfield  {author} {\bibinfo {author} {\bibfnamefont {M.}~\bibnamefont
  {Ausloos}}, \bibinfo {author} {\bibfnamefont {C.}~\bibnamefont {Laurent}},
  \bibinfo {author} {\bibfnamefont {S.~K.}\ \bibnamefont {Patapis}}, \bibinfo
  {author} {\bibfnamefont {C.}~\bibnamefont {Politis}}, \bibinfo {author}
  {\bibfnamefont {H.~L.}\ \bibnamefont {Luo}}, \bibinfo {author} {\bibfnamefont
  {P.~A.}\ \bibnamefont {Godelaine}}, \bibinfo {author} {\bibfnamefont
  {F.}~\bibnamefont {Gillet}}, \bibinfo {author} {\bibfnamefont
  {A.}~\bibnamefont {Dang}}, \ and\ \bibinfo {author} {\bibfnamefont
  {R.}~\bibnamefont {Cloots}},\ }\bibfield  {title} {\enquote {\bibinfo {title}
  {Superconductivity fluctuations in {Bi(Pb)} based granular ceramics
  superconductors: Evidence for fractal behavior},}\ }\href {\doibase
  10.1007/BF01313405} {\bibfield  {journal} {\bibinfo  {journal} {Zeitschrift
  f{\"u}r Physik B Condensed Matter}\ }\textbf {\bibinfo {volume} {83}},\
  \bibinfo {pages} {355--359} (\bibinfo {year} {1991})}\BibitemShut {NoStop}%
\bibitem [{\citenamefont {Bak}(2003)}]{Bak2003}%
  \BibitemOpen
  \bibfield  {author} {\bibinfo {author} {\bibfnamefont {Z.}~\bibnamefont
  {Bak}},\ }\bibfield  {title} {\enquote {\bibinfo {title} {Superconductivity
  in a system of fractional spectral dimension},}\ }\href {\doibase
  10.1103/PhysRevB.68.064511} {\bibfield  {journal} {\bibinfo  {journal} {Phys.
  Rev. B}\ }\textbf {\bibinfo {volume} {68}},\ \bibinfo {pages} {064511}
  (\bibinfo {year} {2003})}\BibitemShut {NoStop}%
\bibitem [{\citenamefont {Kim}, \citenamefont {Rakhimov},\ and\ \citenamefont
  {Yee}(2005)}]{Hyung2005}%
  \BibitemOpen
  \bibfield  {author} {\bibinfo {author} {\bibfnamefont {C.~K.}\ \bibnamefont
  {Kim}}, \bibinfo {author} {\bibfnamefont {A.}~\bibnamefont {Rakhimov}}, \
  and\ \bibinfo {author} {\bibfnamefont {J.~H.}\ \bibnamefont {Yee}},\
  }\bibfield  {title} {\enquote {\bibinfo {title} {{Ginzburg--Landau} theory of
  superconductivity at fractal dimensions},}\ }\href {\doibase
  10.1103/PhysRevB.71.024518} {\bibfield  {journal} {\bibinfo  {journal} {Phys.
  Rev. B}\ }\textbf {\bibinfo {volume} {71}},\ \bibinfo {pages} {024518}
  (\bibinfo {year} {2005})}\BibitemShut {NoStop}%
\bibitem [{\citenamefont {Feigel'man}\ \emph {et~al.}(2010)\citenamefont
  {Feigel'man}, \citenamefont {Ioffe}, \citenamefont {Kravtsov},\ and\
  \citenamefont {Cuevas}}]{Feigelman2010}%
  \BibitemOpen
  \bibfield  {author} {\bibinfo {author} {\bibfnamefont {M.}~\bibnamefont
  {Feigel'man}}, \bibinfo {author} {\bibfnamefont {L.}~\bibnamefont {Ioffe}},
  \bibinfo {author} {\bibfnamefont {V.}~\bibnamefont {Kravtsov}}, \ and\
  \bibinfo {author} {\bibfnamefont {E.}~\bibnamefont {Cuevas}},\ }\bibfield
  {title} {\enquote {\bibinfo {title} {Fractal superconductivity near
  localization threshold},}\ }\href {\doibase
  https://doi.org/10.1016/j.aop.2010.04.001} {\bibfield  {journal} {\bibinfo
  {journal} {Annals of Physics}\ }\textbf {\bibinfo {volume} {325}},\ \bibinfo
  {pages} {1390--1478} (\bibinfo {year} {2010})},\ \bibinfo {note} {july 2010
  Special Issue}\BibitemShut {NoStop}%
\bibitem [{\citenamefont {Sun}\ \emph {et~al.}(2024)\citenamefont {Sun},
  \citenamefont {\ifmmode \check{C}\else \v{C}\fi{}ade\ifmmode~\check{z}\else
  \v{z}\fi{}}, \citenamefont {Yurkevich},\ and\ \citenamefont
  {Andreanov}}]{Meng2024}%
  \BibitemOpen
  \bibfield  {author} {\bibinfo {author} {\bibfnamefont {M.}~\bibnamefont
  {Sun}}, \bibinfo {author} {\bibfnamefont {T.}~\bibnamefont {\ifmmode
  \check{C}\else \v{C}\fi{}ade\ifmmode~\check{z}\else \v{z}\fi{}}}, \bibinfo
  {author} {\bibfnamefont {I.}~\bibnamefont {Yurkevich}}, \ and\ \bibinfo
  {author} {\bibfnamefont {A.}~\bibnamefont {Andreanov}},\ }\bibfield  {title}
  {\enquote {\bibinfo {title} {Enhancement of superconductivity in the
  {Fibonacci} chain},}\ }\href {\doibase 10.1103/PhysRevB.109.134504}
  {\bibfield  {journal} {\bibinfo  {journal} {Phys. Rev. B}\ }\textbf {\bibinfo
  {volume} {109}},\ \bibinfo {pages} {134504} (\bibinfo {year}
  {2024})}\BibitemShut {NoStop}%
\bibitem [{\citenamefont {Poccia}, \citenamefont {Ricci},\ and\ \citenamefont
  {Bianconi}(2011)}]{Poccia2011}%
  \BibitemOpen
  \bibfield  {author} {\bibinfo {author} {\bibfnamefont {N.}~\bibnamefont
  {Poccia}}, \bibinfo {author} {\bibfnamefont {A.}~\bibnamefont {Ricci}}, \
  and\ \bibinfo {author} {\bibfnamefont {A.}~\bibnamefont {Bianconi}},\
  }\bibfield  {title} {\enquote {\bibinfo {title} {Fractal structure favoring
  superconductivity at high temperatures in a stack of membranes near a strain
  quantum critical point},}\ }\href {\doibase 10.1007/s10948-010-1109-x}
  {\bibfield  {journal} {\bibinfo  {journal} {Journal of Superconductivity and
  Novel Magnetism}\ }\textbf {\bibinfo {volume} {24}},\ \bibinfo {pages}
  {1195--1200} (\bibinfo {year} {2011})}\BibitemShut {NoStop}%
\bibitem [{\citenamefont {Meyer}\ \emph {et~al.}(2002)\citenamefont {Meyer},
  \citenamefont {Korshunov}, \citenamefont {Leemann},\ and\ \citenamefont
  {Martinoli}}]{meyer_dimensional_2002}%
  \BibitemOpen
  \bibfield  {author} {\bibinfo {author} {\bibfnamefont {R.}~\bibnamefont
  {Meyer}}, \bibinfo {author} {\bibfnamefont {S.~E.}\ \bibnamefont
  {Korshunov}}, \bibinfo {author} {\bibfnamefont {C.}~\bibnamefont {Leemann}},
  \ and\ \bibinfo {author} {\bibfnamefont {P.}~\bibnamefont {Martinoli}},\
  }\bibfield  {title} {\enquote {\bibinfo {title} {Dimensional crossover and
  hidden incommensurability in {Josephson} junction arrays of periodically
  repeated {Sierpi\'{n}ski} gaskets},}\ }\href {\doibase
  10.1103/PhysRevB.66.104503} {\bibfield  {journal} {\bibinfo  {journal}
  {Physical Review B}\ }\textbf {\bibinfo {volume} {66}},\ \bibinfo {pages}
  {104503} (\bibinfo {year} {2002})}\BibitemShut {NoStop}%
\bibitem [{\citenamefont {Kingni}\ \emph {et~al.}(2019)\citenamefont {Kingni},
  \citenamefont {Kuiate}, \citenamefont {Tamba}, \citenamefont {Monwanou},\
  and\ \citenamefont {Orou}}]{kingni_analysis_2019}%
  \BibitemOpen
  \bibfield  {author} {\bibinfo {author} {\bibfnamefont {S.~T.}\ \bibnamefont
  {Kingni}}, \bibinfo {author} {\bibfnamefont {G.~F.}\ \bibnamefont {Kuiate}},
  \bibinfo {author} {\bibfnamefont {V.~K.}\ \bibnamefont {Tamba}}, \bibinfo
  {author} {\bibfnamefont {A.~V.}\ \bibnamefont {Monwanou}}, \ and\ \bibinfo
  {author} {\bibfnamefont {J.~B.~C.}\ \bibnamefont {Orou}},\ }\bibfield
  {title} {\enquote {\bibinfo {title} {Analysis of a fractal {Josephson}
  junction with unharmonic current--phase relation},}\ }\href {\doibase
  10.1007/s10948-018-4967-2} {\bibfield  {journal} {\bibinfo  {journal}
  {Journal of Superconductivity and Novel Magnetism}\ }\textbf {\bibinfo
  {volume} {32}},\ \bibinfo {pages} {2295--2301} (\bibinfo {year}
  {2019})}\BibitemShut {NoStop}%
\bibitem [{\citenamefont {Manna}, \citenamefont {Das},\ and\ \citenamefont
  {Roy}(2022)}]{manna2022noncrystalline}%
  \BibitemOpen
  \bibfield  {author} {\bibinfo {author} {\bibfnamefont {S.}~\bibnamefont
  {Manna}}, \bibinfo {author} {\bibfnamefont {S.~K.}\ \bibnamefont {Das}}, \
  and\ \bibinfo {author} {\bibfnamefont {B.}~\bibnamefont {Roy}},\ }\href@noop
  {} {\enquote {\bibinfo {title} {Noncrystalline topological
  superconductors},}\ } (\bibinfo {year} {2022}),\ \Eprint
  {http://arxiv.org/abs/2207.02203} {arXiv:2207.02203} \BibitemShut {NoStop}%
\bibitem [{\citenamefont {Iliasov}, \citenamefont {Katsnelson},\ and\
  \citenamefont {Bagrov}(2024)}]{iliasov2024}%
  \BibitemOpen
  \bibfield  {author} {\bibinfo {author} {\bibfnamefont {A.~A.}\ \bibnamefont
  {Iliasov}}, \bibinfo {author} {\bibfnamefont {M.~I.}\ \bibnamefont
  {Katsnelson}}, \ and\ \bibinfo {author} {\bibfnamefont {A.~A.}\ \bibnamefont
  {Bagrov}},\ }\href {http://arxiv.org/abs/2310.11497} {\enquote {\bibinfo
  {title} {Strong enhancement of superconductivity on finitely ramified fractal
  lattices},}\ } (\bibinfo {year} {2024}),\ \Eprint
  {http://arxiv.org/abs/2310.11497} {arXiv:2310.11497} \BibitemShut {NoStop}%
\bibitem [{\citenamefont {Josephson}(1962)}]{josephson1962}%
  \BibitemOpen
  \bibfield  {author} {\bibinfo {author} {\bibfnamefont {B.}~\bibnamefont
  {Josephson}},\ }\bibfield  {title} {\enquote {\bibinfo {title} {Possible new
  effects in superconductive tunnelling},}\ }\href {\doibase
  10.1016/0031-9163(62)91369-0} {\bibfield  {journal} {\bibinfo  {journal}
  {Physics Letters}\ }\textbf {\bibinfo {volume} {1}},\ \bibinfo {pages}
  {251--253} (\bibinfo {year} {1962})}\BibitemShut {NoStop}%
\bibitem [{\citenamefont {Golubov}, \citenamefont {Kupriyanov},\ and\
  \citenamefont {Il’ichev}(2004)}]{golubov2004a}%
  \BibitemOpen
  \bibfield  {author} {\bibinfo {author} {\bibfnamefont {A.~A.}\ \bibnamefont
  {Golubov}}, \bibinfo {author} {\bibfnamefont {M.~Y.}\ \bibnamefont
  {Kupriyanov}}, \ and\ \bibinfo {author} {\bibfnamefont {E.}~\bibnamefont
  {Il’ichev}},\ }\bibfield  {title} {\enquote {\bibinfo {title} {The
  current--phase relation in {Josephson} junctions},}\ }\href {\doibase
  10.1103/revmodphys.76.411} {\bibfield  {journal} {\bibinfo  {journal}
  {Reviews of Modern Physics}\ }\textbf {\bibinfo {volume} {76}},\ \bibinfo
  {pages} {411--469} (\bibinfo {year} {2004})}\BibitemShut {NoStop}%
\bibitem [{\citenamefont {Kruchinin}\ \emph {et~al.}(2005)\citenamefont
  {Kruchinin}, \citenamefont {Klepikov}, \citenamefont {Novikov},\ and\
  \citenamefont {Kruchinin}}]{kruchinin_nonlinear_2005}%
  \BibitemOpen
  \bibfield  {author} {\bibinfo {author} {\bibfnamefont {S.~P.}\ \bibnamefont
  {Kruchinin}}, \bibinfo {author} {\bibfnamefont {V.~F.}\ \bibnamefont
  {Klepikov}}, \bibinfo {author} {\bibfnamefont {V.~E.}\ \bibnamefont
  {Novikov}}, \ and\ \bibinfo {author} {\bibfnamefont {D.~S.}\ \bibnamefont
  {Kruchinin}},\ }\bibfield  {title} {\enquote {\bibinfo {title} {Nonlinear
  current oscillations in the fractal {Josephson} junction},}\ }\href@noop {}
  {\bibfield  {journal} {\bibinfo  {journal} {Materials Science Poland}\
  }\textbf {\bibinfo {volume} {23}},\ \bibinfo {pages} {1009} (\bibinfo {year}
  {2005})}\BibitemShut {NoStop}%
\bibitem [{\citenamefont {Bardeen}, \citenamefont {Cooper},\ and\ \citenamefont
  {Schrieffer}(1957)}]{bardeen1957}%
  \BibitemOpen
  \bibfield  {author} {\bibinfo {author} {\bibfnamefont {J.}~\bibnamefont
  {Bardeen}}, \bibinfo {author} {\bibfnamefont {L.~N.}\ \bibnamefont {Cooper}},
  \ and\ \bibinfo {author} {\bibfnamefont {J.~R.}\ \bibnamefont {Schrieffer}},\
  }\bibfield  {title} {\enquote {\bibinfo {title} {Theory of
  superconductivity},}\ }\href {\doibase 10.1103/physrev.108.1175} {\bibfield
  {journal} {\bibinfo  {journal} {Physical Review}\ }\textbf {\bibinfo {volume}
  {108}},\ \bibinfo {pages} {1175--1204} (\bibinfo {year} {1957})}\BibitemShut
  {NoStop}%
\bibitem [{\citenamefont {Dynes}\ and\ \citenamefont
  {Fulton}(1971)}]{dynes1971a}%
  \BibitemOpen
  \bibfield  {author} {\bibinfo {author} {\bibfnamefont {R.~C.}\ \bibnamefont
  {Dynes}}\ and\ \bibinfo {author} {\bibfnamefont {T.~A.}\ \bibnamefont
  {Fulton}},\ }\bibfield  {title} {\enquote {\bibinfo {title} {Supercurrent
  density distribution in {Josephson} junctions},}\ }\href {\doibase
  10.1103/physrevb.3.3015} {\bibfield  {journal} {\bibinfo  {journal} {Physical
  Review B}\ }\textbf {\bibinfo {volume} {3}},\ \bibinfo {pages} {3015--3023}
  (\bibinfo {year} {1971})}\BibitemShut {NoStop}%
\bibitem [{\citenamefont {Cohen}(1995)}]{fractal_antennae}%
  \BibitemOpen
  \bibfield  {author} {\bibinfo {author} {\bibfnamefont {N.}~\bibnamefont
  {Cohen}},\ }\href {https://patents.google.com/patent/US6452553} {\emph
  {\bibinfo {title} {Fractal antennas and fractal resonators}}}\ (\bibinfo
  {publisher} {{U.S. Patent 6452553}},\ \bibinfo {year} {1995})\BibitemShut
  {NoStop}%
\bibitem [{\citenamefont {Cohen}(1997)}]{Cohen_fractal}%
  \BibitemOpen
  \bibfield  {author} {\bibinfo {author} {\bibfnamefont {N.}~\bibnamefont
  {Cohen}},\ }\bibfield  {title} {\enquote {\bibinfo {title} {Fractal antenna
  applications in wireless telecommunications},}\ }in\ \href {\doibase
  10.1109/EIF.1997.605374} {\emph {\bibinfo {booktitle} {Professional Program
  Proceedings. Electronic Industries Forum of New England}}}\ (\bibinfo {year}
  {1997})\ pp.\ \bibinfo {pages} {43--49}\BibitemShut {NoStop}%
\bibitem [{\citenamefont {Werner}\ and\ \citenamefont
  {Ganguly}(2003)}]{Werner_fractal}%
  \BibitemOpen
  \bibfield  {author} {\bibinfo {author} {\bibfnamefont {D.}~\bibnamefont
  {Werner}}\ and\ \bibinfo {author} {\bibfnamefont {S.}~\bibnamefont
  {Ganguly}},\ }\bibfield  {title} {\enquote {\bibinfo {title} {An overview of
  fractal antenna engineering research},}\ }\href {\doibase
  10.1109/MAP.2003.1189650} {\bibfield  {journal} {\bibinfo  {journal} {IEEE
  Antennas and Propagation Magazine}\ }\textbf {\bibinfo {volume} {45}},\
  \bibinfo {pages} {38--57} (\bibinfo {year} {2003})}\BibitemShut {NoStop}%
\bibitem [{\citenamefont {Hori}\ \emph {et~al.}(2024)\citenamefont {Hori},
  \citenamefont {Sugimoto}, \citenamefont {Tohyama},\ and\ \citenamefont
  {Tanaka}}]{hori2024}%
  \BibitemOpen
  \bibfield  {author} {\bibinfo {author} {\bibfnamefont {M.}~\bibnamefont
  {Hori}}, \bibinfo {author} {\bibfnamefont {T.}~\bibnamefont {Sugimoto}},
  \bibinfo {author} {\bibfnamefont {T.}~\bibnamefont {Tohyama}}, \ and\
  \bibinfo {author} {\bibfnamefont {K.}~\bibnamefont {Tanaka}},\ }\href
  {http://arxiv.org/abs/2401.06355} {\enquote {\bibinfo {title}
  {Self-consistent study of topological superconductivity in two-dimensional
  quasicrystals},}\ } (\bibinfo {year} {2024}),\ \Eprint
  {http://arxiv.org/abs/2401.06355} {arXiv:2401.06355} \BibitemShut {NoStop}%
\end{thebibliography}%
\end{document}